\documentclass[12pt]{iopart}

\usepackage{graphicx}
\begin{document}

\title[ Amorphous silicon on textured and non-textured surface]{Analysis of growth of silicon thin films on textured and non-textured surface}

\author{S. M. Iftiquar$^{1}$, S. N. Riaz$^{1,2}$, S. Mahapatra$^{3}$}

\address{$^{1}$College of information and Communications Engineering, Sunkyunkwan University, Suwon, South Korea }
\address{$^{2}$School of Material Science and Nanotechnology, Jadavpur University, 188-Raja Subodh Chandra  Mallick Road, Jadavpur, Kolkata-700 032, West Bengal, India}
\address{$^{3}$Department of Energy, Tezpur University, Napaam, Tezpur 784028, Assam, India }
\ead{smiftiquar@gmail.com}

\begin{abstract}

Hydrogenated amorphous silicon alloy films are generally deposited by radio frequency plasma enhanced chemical vapor deposition (RF PECVD) technique on various types of substrates. Generally it is assumed that film quality remains unchanged when deposited on textured or non-textured substrates. Here we analyzed the difference in growth of thin film silicon layers when deposited in a textured and a non-textured surface. In this investigation characteristics of two solar cells were compared, where one cell was prepared on a textured surface ( Cell-A) while the other prepared on a non-textured surface (Cell-B). Defect analysis of the devices were carried out by simulation and device modeling. It shows that the intrinsic film deposited on a textured surface was more defective ($2.4\times 10^{17}$ cm$^{-3}$) than that deposited on a flat surface ($3.2\times 10^{16}$ cm$^{-3}$). Although the primary differences in these two cells were thickness of the active layer and nature of surface texturing, the simulation results show that thin film deposited on a textured surface may acquire an increased defect density than that deposited on a flat surface. Lower effective flux density of $SiH_{3}$ precursors on the textured surface can be one of the reasons for higher defect density in the film deposited on textured surface. An Improved light coupling can be achieved by using a thinner doped window layer. By changing the thickness from 15 nm to 3 nm, the short circuit current density increased from 16.4 mA/cm$^{2}$ to 20.96 mA/cm$^{2}$ and efficiency increased from $9.4\%$ to $12.32\%$. 

\vspace{2pc}
\noindent{\bf keywords}: defect generation; textured surface; non-textured surface; plasma deposition; numerical simulation; defect density; window layer  

\end{abstract}
 
\maketitle

\section{Introduction}

Thin film silicon solar cell has long been investigated for photovoltaic conversion of sun light. There have been various successes in the past where high efficiency of the devices were reported. Single junction thin film silicon solar cell is composed of an active layer or a single set of p-i-n structure. Power conversion efficiency (PCE) of around 10$\%$ were reported [1-3]. Tandem solar cell, consisting of two active layers or two p-i-n type sub-cells, was reported to have efficiency higher than the single junction device [4]. However the efficiency always remained lower than that of crystalline silicon solar cell or silicon hetero junction solar cell (HJSC) with thin intrinsic layer. One of the interesting characteristics of the HJSC is that it can generate high current density under short circuit condition ($J_{sc}$), but its output voltage remained lower than that of thin film silicon solar cell [5, 6]. Work has been going on with a tandem solar cell having top sub-cell as thin film silicon [7, 8]. It shows higher efficiency than that of the thin film silicon but lower current density of the top sub-cell leads to a lower overall current density because current density of a two terminal tandem cell is limited by the lowest current generating sub-cell.

Therefore raising current density of the top sub-cell is desirable for high efficiency solar cell. Various attempts have been made to raise the current density, like raising thickness of the top sub-cell and using different materials are two of the possibilities [8]. But the drawback of raising thickness is that it inadvertently reduces available light for the bottom sub-cell. In most of the cases this approach does not lead to a higher efficiency.

\subsection{Light trapping with textured surface}

Introducing a light trapping scheme is another approach to raise the current density [9]. The method involves more light to pass though the absorber or active layer, thereby enhancing absorption of more light. Texturing front surface can significantly reduce optical reflection thereby capture larger part of  the incident radiation or enhances the light coupling into the cell, which helps increasing the current density [10-12].

Usually the size of surface texturing is larger than visible wavelength of light and is of several micrometer for each structure. One approach is to deposit the absorber layer is deposited directly on such a surface, so the deposited films take shape of the surface texture.
   
But this approach of fabricating solar cell (on a textured surface) may not be free from its limitation either. Because the surface is uneven, the thin films prepared by plasma deposition become uneven as well, leading to texture induced surface defects of the deposited films [13].

There have been some reports on the texture induced surface defects [14-16]. The defects can be of various types, one of them is physical where the deposited film is not as uniform as expected, or the defect can be electronic, where, even if the film is physically uniform, its defect density increases due to the surface texturing. In a well controlled plasma deposition the physical defect can be considered as minimum, but its electronic defect may increase. For this reason, films deposited on a textured surface may be more defective, and in a photovoltaic (PV) device its effect can become obvious in the form of low open circuit voltage, fill factor as well as current density. 

Based on this assumption we selected two reported solar cells, one was deposited on a flat surface while the other on a textured surface. In order to obtain material properties, AFORS-HET simulation program [17] was used. 

\subsection{Simulation approach}

In the AFORS-HET simulation program, explicit details of the device structure and material properties are required. Although the device structure is available in the literatures, yet the material characteristics (like defect density, carrier mobility etc) for each of the layers had to be adapted partly from other published literatures and partly from the simulation. 

The J-V characteristic curves obtained from the simulation depends on the parameters chosen. In other words, by varying the  electronic properties of the layers and simulating the J-V curve, we obtained a large number of the characteristic curves. Out of them, a few were very close to that of Cell-A and Cell-B. 

The simulated curves that were close to the actual J-V curves, were chosen as the most probable solutions. The material properties of the chosen devices were used to compare the materials that were deposited on the flat and textured surface.  A comparison between the cells was used to understand the differences in electronic characteristics of the layers deposited on the flat and textured surfaces. Furthermore, with the help of the simulation, an optimum device structure was obtained for a maximum PCE in each of the two types of cells. 

Furthermore, a diode equivalent model of solar cell was also used to extract the related equivalent parameters. Reverse saturation current density ($I_{0}$) gives an indication of defects in the device; higher $I_{0}$ indicates more defective layer.   

\section{Device Description}

In order to investigate the texture induced electronic defects, we preferred two solar cells as shown in Fig. 1(a), (b). A diode equivalent circuit model of solar cell is shown in Fig. 1(c). Details of these two devices are in references [3, 12]. The cell-A [12] was prepared on a textured cSi (Fig. 1(a)) and the other one (Fig 1(b)) was prepared on a TCO coated glass substrate (Cell-B)[3]. Both these cells are of similar types with intrinsic amorphous silicon active layer, p-type silicon oxide and n-type amorphous silicon as doped layers. Both of them have p-type layer as front window layer with Ag/Al at the back side.

The structure of the Cell-A is (on textured surface): p(20nm)/i(225nm)/n(25nm) .
The structure of Cell-B is  (on flat surface): p(15nm)/i(450nm)/n(25nm).

The current voltage characteristics of the these two cells under AM1.5G insolation are shown in Fig. 2(a),(b) while the characteristic parameters that were extracted from the J-V curves, are shown within the respective figures as legend. The simulated J-V curves are shown as continuous lines, while the marked data points are from the reported cell data.

\begin{figure}
	\centering
	\includegraphics[width=6cm]{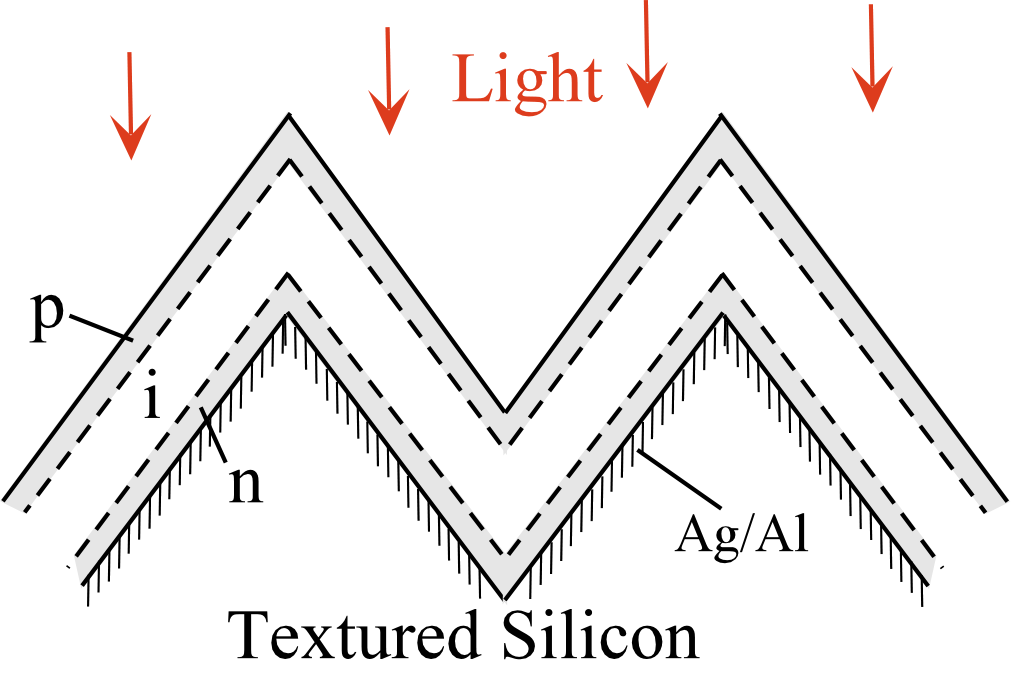} %\label{a} 
	\includegraphics[width=8cm]{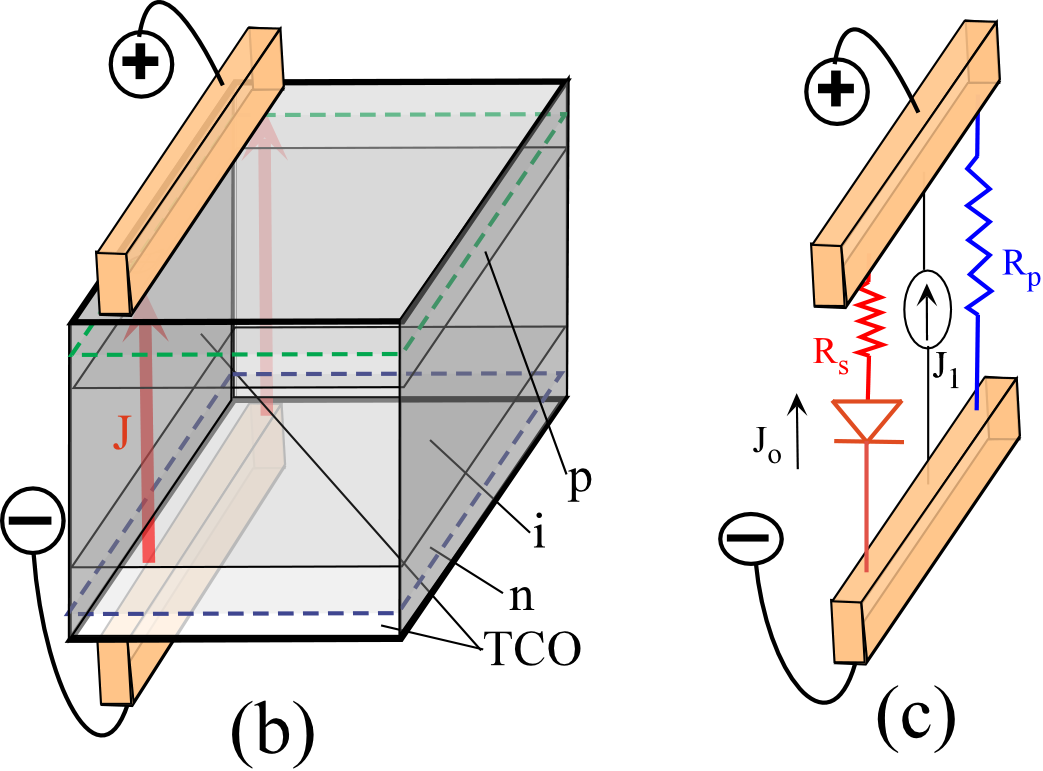}    %\end{figure}
	\caption[(a)]{(a) Schematic diagram of the two solar cells used in the investigation. Cell-A on a textured surface, (b) Cell-B on a flat surface, (c) diode equivalent circuit of a solar cell. }
	\label{fig:fig1}
\end{figure}

%(a)

%Fig. 1. 

\begin{figure}
	\centering
	\includegraphics[width=7cm]{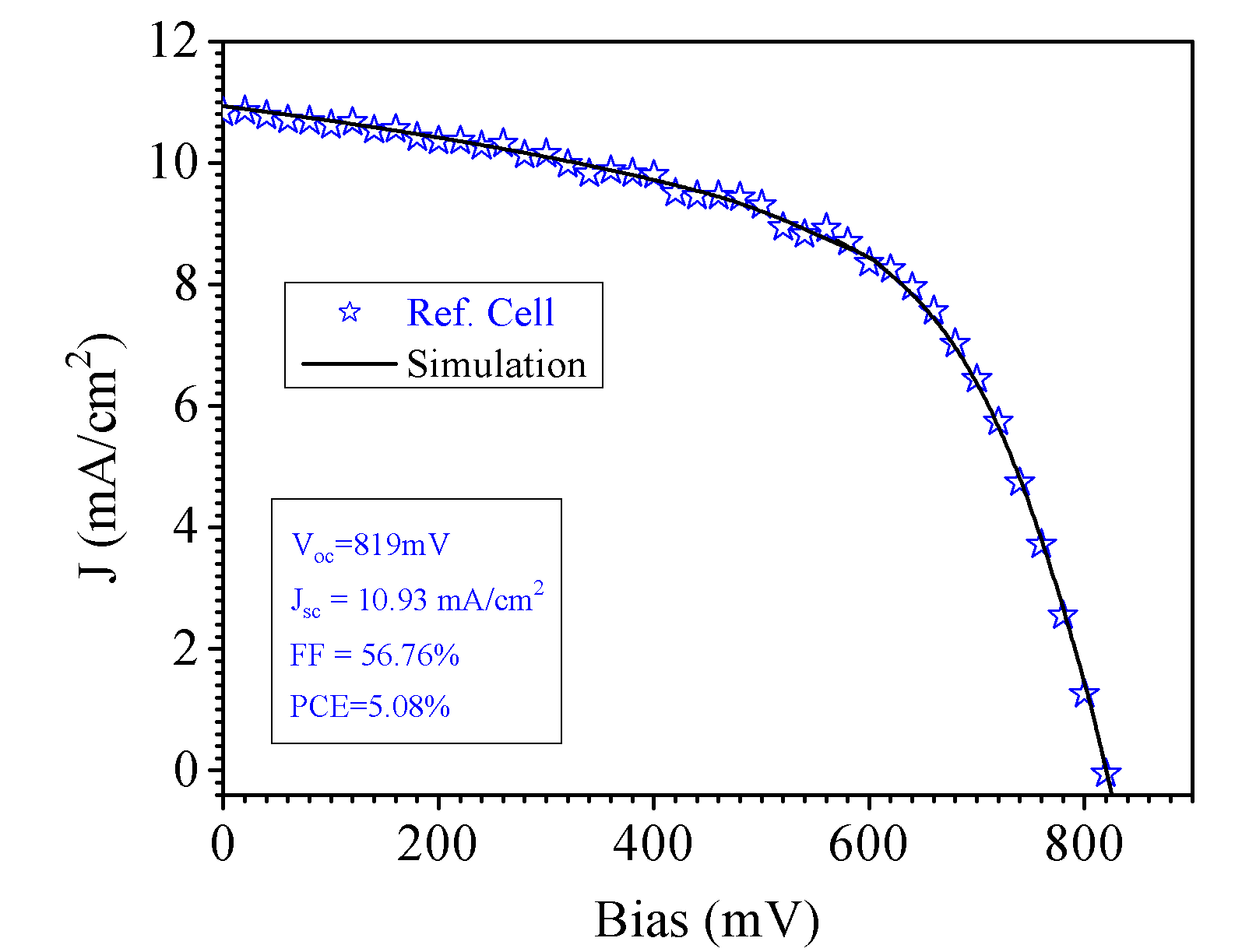} %\label{a} 
	\includegraphics[width=7cm]{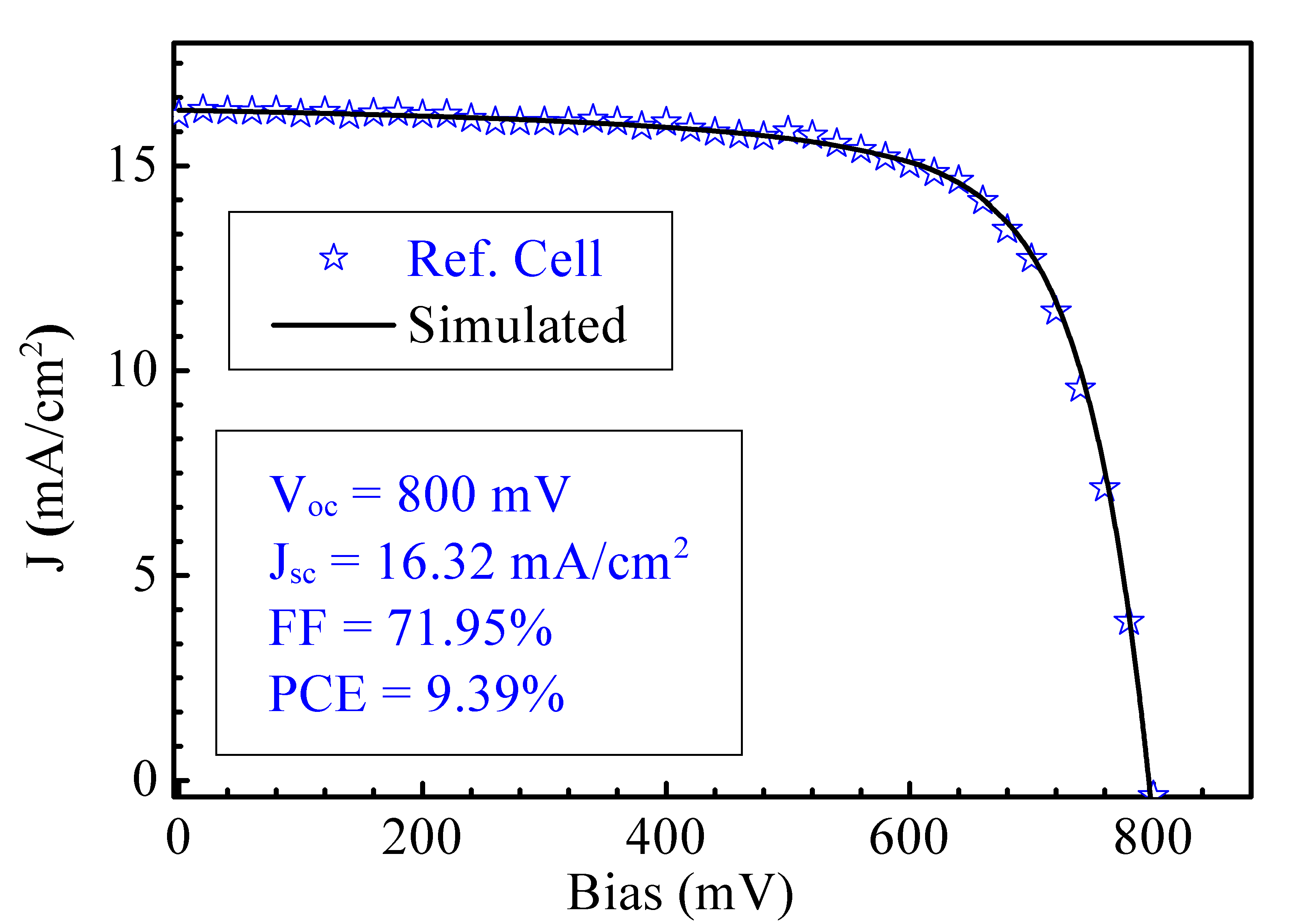}    %\end{figure}
	\caption[(a)]{(a) Current density voltage (J-V) characteristic curves of the two solar  cell where the ‘star ’ indicates the data points from real cell while the continuous lines are simulated J-V curves that match closely to these curves. (a) Cell-A [12], (b) Cell-B[3] }
	\label{fig:fig2}
\end{figure}

%Table I. 

\begin{table}
	\caption{\label{jlab1} Simulation parameters for the Cell-A, the starting parameters for simulation of the J-V characteristics of the Cell-B and then the final parameters of the layers for which the J-V characteristics of the real and simulated characteristics are very close to each other. Here $N_{c}$ is conduction band density of states (DOS), $N_{v}$ is valence band DOS, $N_{trap}$ is sum of donor and acceptor DOS, $\mu_{e}$ is electron mobility, $\mu_{h}$ is hole mobility, $N_{d}$ is donor density, $N_{a}$ is acceptor density.}

	\begin{center}
		\footnotesize
		\begin{tabular}{@{}lllll}
			\br
			&  &   & Initial & Final   \\
			\mr
			& Layer & Cell-A [12]           & Cell-B & Cell-B  \\
			& name  &                       &        &               \\
			Layer      & p     & 20                    & 15     &  15   \\
			Thickness  & i     & 225                   & 450    &  450   \\
			d (nm)     & n     & 25                    & 25     &  25   \\
			\mr
			$N_{c}$    &  p    & $3.0\times 10^{20}$   & $3.0\times 10^{20}$  &  $8.5\times 10^{20}$   \\
			(cm$^{-3}$)& i     &  $1.1\times 10^{19}$  & $3.0\times 10^{20}$  & $8.5\times 10^{20}$    \\
			&  n    &  $1.0\times 10^{20}$  & $3.0\times 10^{20}$  &  $8.5\times 10^{20}$   \\
			\mr
			$N_{v}$    &  p    & $3.0\times 10^{20}$   & $8.5\times 10^{20}$  &  $8.5\times 10^{20}$   \\
			(cm$^{-3}$)& i     &  $1.1\times 10^{19}$  & $3.0\times 10^{20}$  & $8.5\times 10^{20}$    \\
			&  n    &  $1.0\times 10^{20}$  & $3.0\times 10^{20}$  &  $8.5\times 10^{20}$   \\
			\mr
			$N_{trap}$ (cm$^{-3}$)&  p    & $1.0\times 10^{18}$   & $3.2\times 10^{16}$  &  $3.2\times 10^{16}$   \\
			donor,	& i     &  $2.4\times 10^{17}$  & $3.4\times 10^{17}$  & $3.2\times 10^{16}$    \\
			acceptor	&  n    &  $1.4\times 10^{19}$  & $3.2\times 10^{16}$  &  $3.2\times 10^{16}$   \\
			\mr
			$\mu_{e}$       & p & 1.5 & 0.1 &  0.1  \\
			(cm$^{2}$/V.s)  & i & 7.0 & 7.0 & 1.0   \\
			(cm$^{-3}$ )    & n & 0.5 & 3.0 & 3.0    \\
			\mr
			$\mu_{h}$       & p & 0.37 & 3.0 &  3.0  \\
			(cm$^{2}$/V.s)  & i & 0.18 & 0.18 & 0.05   \\
			(cm$^{-3}$ )    & n & 0.1 & 0.1 & 0.1    \\
			\mr
			$N_{d}$ cm$^{-3}$ & n & $7.5\times 10^{19}$   & $1.0\times 10^{19}$ & $1.0\times 10^{19}$   \\
			\mr
			$N_{a}$ cm$^{-3}$ & p & $6.9\times 10^{19}$   & $6.9\times 10^{19}$ & $1.0\times 10^{19}$   \\
			\mr
			Surface&    & Textured & Flat & Flat   \\
			\br
		\end{tabular}\\
	\end{center}
\end{table}

%Table II. 

\begin{table}
	\caption{\label{jlab2} Variation of parameters}
	
	\begin{center}
		\footnotesize
		\begin{tabular}{@{}llll}
			\br
			$N_{c}, N_{v}$ (cm$^{-3}$) & $N_{trap}$(cm$^{-3}$)  & $\mu_{e}, \mu_{h}$ (cm$^{2}$.V.s)  & $\mu_{e}, \mu_{h}$ (cm$^{2}$.V.s)    \\
			(p, i, n layer) & (i layer) & (p, n layer)  & (i layer)    \\
			\mr
			
			$4.0\times 10^{20}$      & $1.0\times 10^{17}$   & 3,\ \ \ 0.1  & 6,\  0.1        \\
			\mr
			$5.0\times 10^{20}$      & $9.0\times 10^{16}$   &   & 5,\ \ \   0.09        \\
			\mr
			$6.0\times 10^{20}$      & $8.0\times 10^{16}$   &   & 4,\ \ \  0.08        \\
			\mr
			$7.0\times 10^{20}$      & $7.0\times 10^{16}$   &   & 3,\ \ \  0.07        \\
			\mr
			$8.0\times 10^{20}$      & $6.0\times 10^{16}$   &   & 2,\ \ \  0.06        \\
			\mr
			$8.5\times 10^{20}$      & $5.0\times 10^{16}$   &   & 1,\  \ \ 0.05        \\
			\mr
			$9.0\times 10^{20}$      & $4.0\times 10^{16}$   &   &         \\
			\mr
			& $3.0\times 10^{16}$   &   &          \\
			\mr
			& $2.0\times 10^{16}$   &   &          \\
			\mr
			& $1.6\times 10^{16}$   &   &          \\
			\mr
			& $1.0\times 10^{16}$   &   &          \\
			\br
		\end{tabular}\\
	\end{center}
\end{table}

\section{Simulation}

The simulated J-V characteristic curves were obtained with the help of AFORS-HET simulation program [17]. The starting simulation parameters were adapted from that of the Cell-A [12], with a small modifications based partly of the physical dimension of the device and partly on assumption. We assumed that for the Cell-B the density of states of the valence and conduction band states will be higher than that in Cell-A, its defect density and carrier mobility will be lower because performance parameters of Cell-B is better than that of Cell-A. Table I shows the simulation parameters for the Cell-A, the starting parameters for simulation of the J-V characteristics of the Cell-B and then the final parameters of the layers for which the J-V characteristics of the real and simulated characteristics are very close to each other.

The step by step parameter variations for the simulation is shown in Table II. We assumed that the films deposited on the flat surface will be denser than that of the textured surface, so the extended density of states will be higher. The density of electronic trap is also expected to be lower for the films on the flat surface. As the films on the flat surface is expected to have lower defect density, so defect mediated conduction is also expected to be lower, or carrier mobility will remain lower. Based on these major and basic characteristics the simulation parameters were changed, and shown in Table II. However, towards the end and for best match of the J-V characteristic curves, a fine tuning of the electronic parameters were made.

 \section{Results and Discussions}

The various J-V characteristic curves that we obtained are shown in Fig. 3. It shows that with the progress of these variations, the open circuit voltage ($V_{oc}$) and $J_{sc}$ of the cells change and become very close to that of the Cell-B. The reference J-V characteristic curve and the closest J-V curves are shown in Fig. 2(b). In Fig. 3 the reference J-V curve is shown as dotted line, while the final simulation is shown as the closest curve to it, with $N_{c}, N_{v} = 8.5\times 10^{20}$. 

\begin{figure}
	\centering
	\includegraphics[width=15cm]{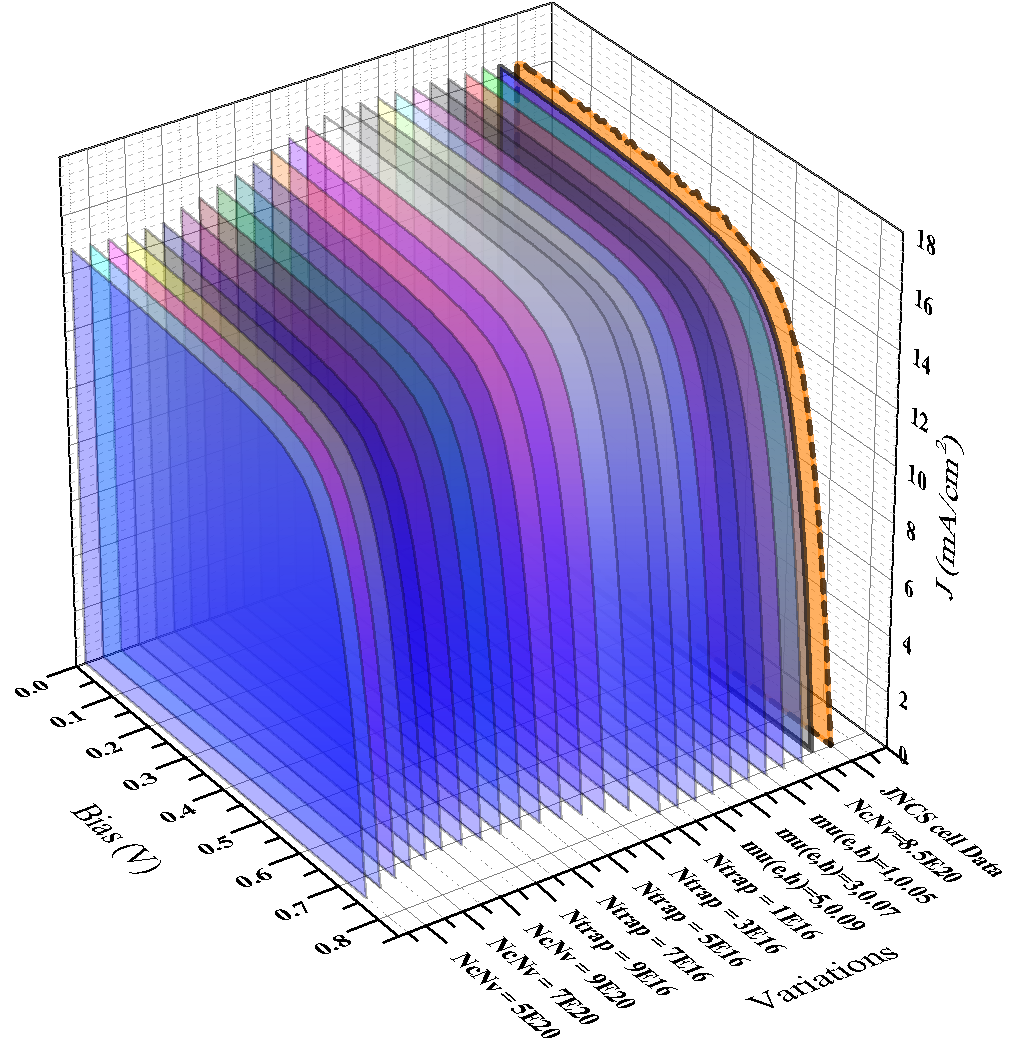} %\label{a} 
	\caption[(a)]{J-V characteristic curves of the simulated solar cells from the starting set of parameters to the final device characteristic. Here the ‘Ref. Cell’ corresponds to the real cell (Cell-B) and the curve immediately close to this curve is the best matched one.}
	\label{fig:fig3}
\end{figure}

%Fig. 3.  

During the simulation, three primary parameters were varied along with a fine tuning for the best possible match. These primary parameters are valence and conduction band density of states (DOS), trap densities and carrier mobility. In this process the variation in device parameters are plotted and shown in Fig.4. Here, three major regions are also indicated. The region (1) is where the valence and conduction band DOS were increased, in the region (2) the defect densities ($N_{trap}$) of the active layer was reduced while in region (3) carrier mobility (electron home mobilities $\mu_{e}$, $\mu_{h}$) were reduced. In the final region (4) fine tuning of the parameters were made to arrive very close to the characteristics of the Cell-B. The Fig. 4 shows that with the increase in the DOS, the $V_{oc}$, $J_{sc}$, FF, PCE, PmaxV, PmaxJ decreases. On the other hand with a reduction in the $N_{trap}$ all these parameters increase. In the region (3) when the electron hole mobility was gradually reduced, $V_{oc}$ and PmaxV increased whereas the other parameters ($J_{sc}$, FF, PCE, PmaxJ ) reduced. In these figure the “star” symbols indicate the parameters for the physical cell (Cell-B). 

\begin{figure}
	\centering
	\includegraphics[width=15cm]{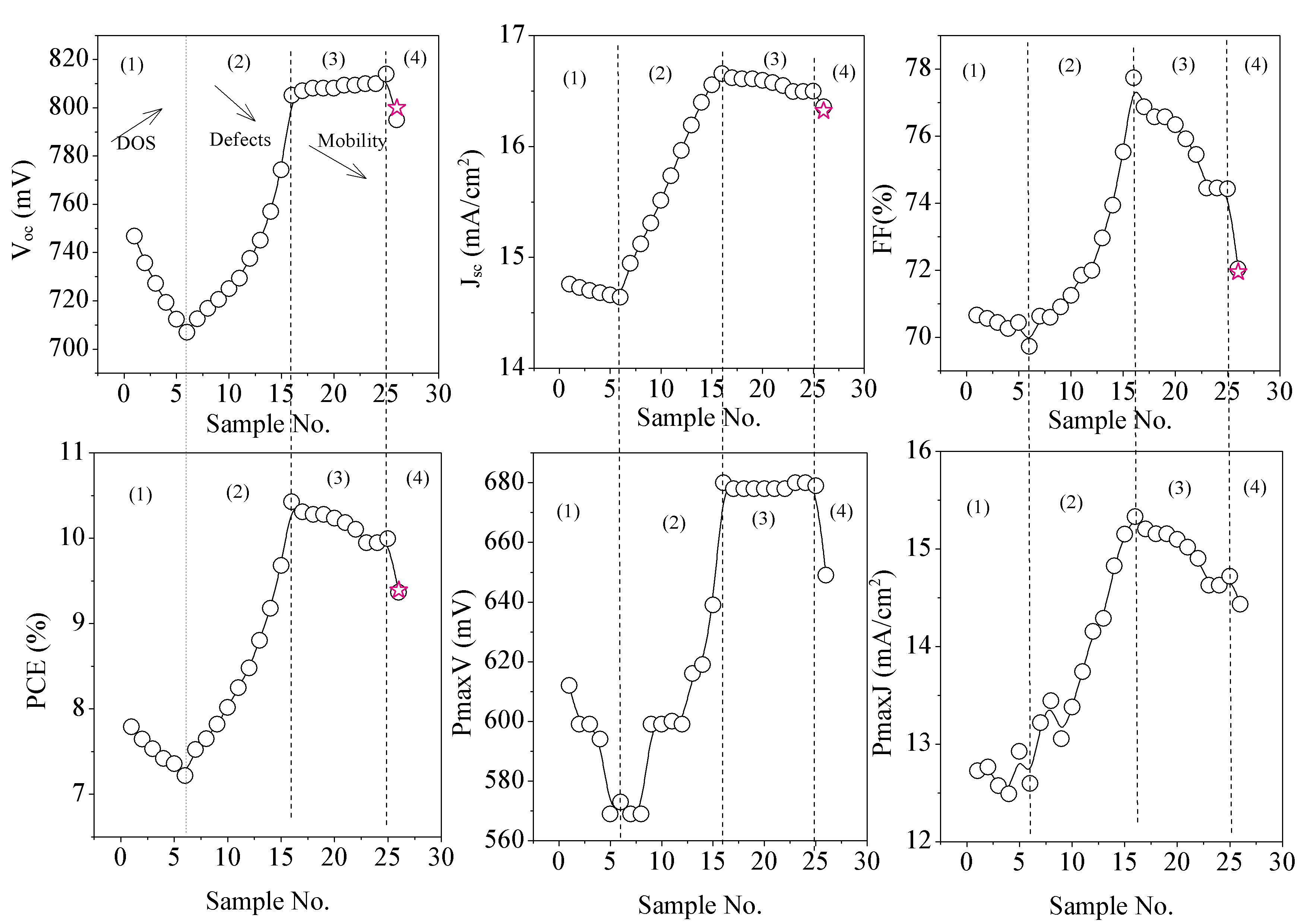} %\label{a} 
	\caption[(a)]{Parameters extracted from the J-V characteristic of Fig. 3. Here $V_{oc}$ is open circuit voltage, $J_{sc}$ is short circuit current density, FF is fill factor, PCE is power conversion efficiency, PmaxV and PmaxJ are the voltage and current density respectively, at the maximum power point. }
	\label{fig:fig4}
\end{figure}

%Fig. 4. 

A single junction solar cell can be represented by a diode equivalent circuit with series ($R_{s}$), shunt resistances ($R_{p}$), a diode and a current generator, as shown in Fig. 1(c) and equation(1). The equation for the diode equivalent circuit is as given in the equation [18, 19].

\begin{figure}
	\centering
	\includegraphics[width=15cm]{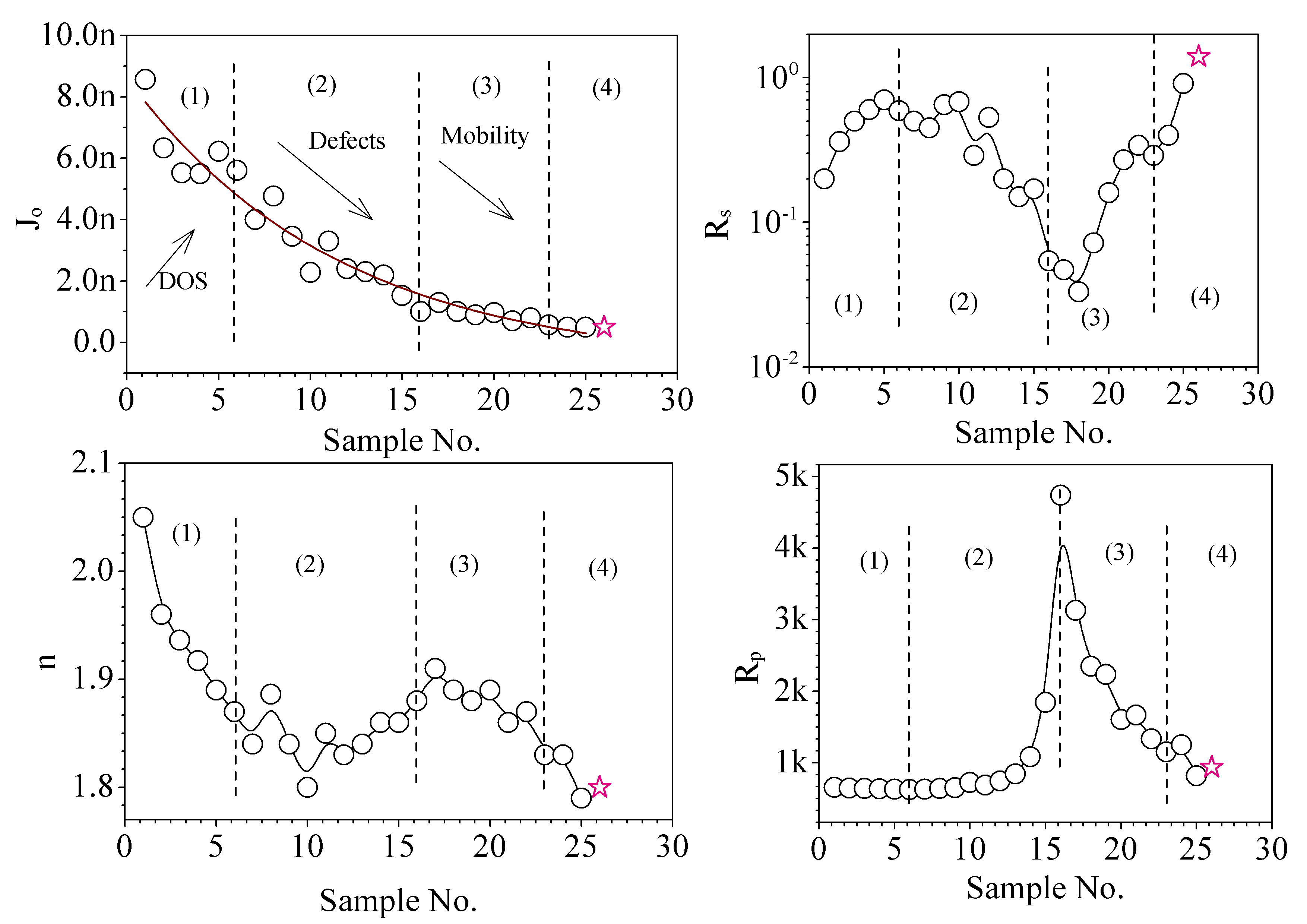} %\label{a} 
	\caption[(a)]{Extracted diode parameters from the J-V characteristic curves of Fig. 3. $J_{0}$ is reverse saturation current density (in A/cm$^{2}$), $R_{s}$ is series resistance (in $\Omega$.cm$^{2}$), n is diode ideality factor, $R_{p}$ is shunt resistance (in $\Omega$.cm$^{2}$) }
	\label{fig:fig1}
\end{figure}

%(1)
%
%
%Fig. 5. .

\begin{equation}\label{key}
	J = J_{sco} - J_{0} \left[e^{\left(\frac{qV+qJR_{s}}{nkT}  \right)}  -1 \right] -\frac{V+JR_{s}}{R_{p}}
\end{equation}

Here $J_{sco}$ is strength of current source which is also considered equal to short circuit current density ($J_{sc}$), $J_{0}$ is reverse saturation current density, n is diode ideality factor, q is charge of electron. These diode parameters were evaluated from the J-V curves, and are shown in Fig. 5. 

\subsection{Reverse saturation current density, $J_{0}$ }
The Fig. 5 shows that $J_{0}$ (reverse saturation current density) steadily decreases in all the four regions, with increasing DOS, reducing defects, and reducing carrier mobility. It is especially to be noticed that the $J_{0}$ reduces with reduced defect density. It is known that higher $J_{0}$ implies more defective material and vice versa.

\subsection{Series resistance, $R_{s}$ }

The series resistance provides dissipation for current flow. With increase in DOS, the material density is expected to increase leading to an increase in specific resistivity. So in the region (1) in increases with the increase in DOS. The $R_{s}$ can partly be due to scattering of the carriers and defect centers. With a reduction in defects in region (2) the reduction in the $R_{s}$ indicates a possible contribution of carrier scattering from the defect centers, thereby a reduction in defect density leads to a reduction in the $R_{s}$. Carrier mobility is inversely related to resistance, that is why when the carrier mobility was reduced the $R_{s}$ was observed to increase.  

\subsection{Shunt resistance, $R_{p}$ }

The shunt resistance is related to recombination of the photo-generated charge carriers. In a lumped circuit model it indicates a part of the photo generated current is lost by shunting through the model $R_{p}$. Higher recombination is indicated by lower $R_{p}$ and vice versa. With an increased DOS, the $R_{p}$ did not show significant variation and remained close to 750 $\Omega$.cm$^{2}$, may be because of the presence of higher defect density. So when the defect density was reduced the $R_{p}$ started increasing until reaching about 5 k$\Omega$.cm$^{2}$. With a reduced carrier mobility, more of the photo generated carriers are expected the be lost by recombination. A similar indication comes from the reduction in the $R_{p}$ with a reduction in the mobility. 

\subsection{Diode ideality factor, $n$ }

The diode ideality factor (n) can be considered as a shielding or screening factor for the electrical bias (V). If carrier density increases, it can raise the shielding effect. If there are increased carrier recombination then average density of carriers is expected to fall, leading to a fall in n. The results show that when the DOS was increased the n reduced, indicating that the shielding effect can also be higher at a higher DOS. Based on a similar reason, when the $N_{trap}$ was reduced the n started increasing and with the reduced mobility the n reduced.

\begin{figure}
	\centering
	\includegraphics[width=15cm]{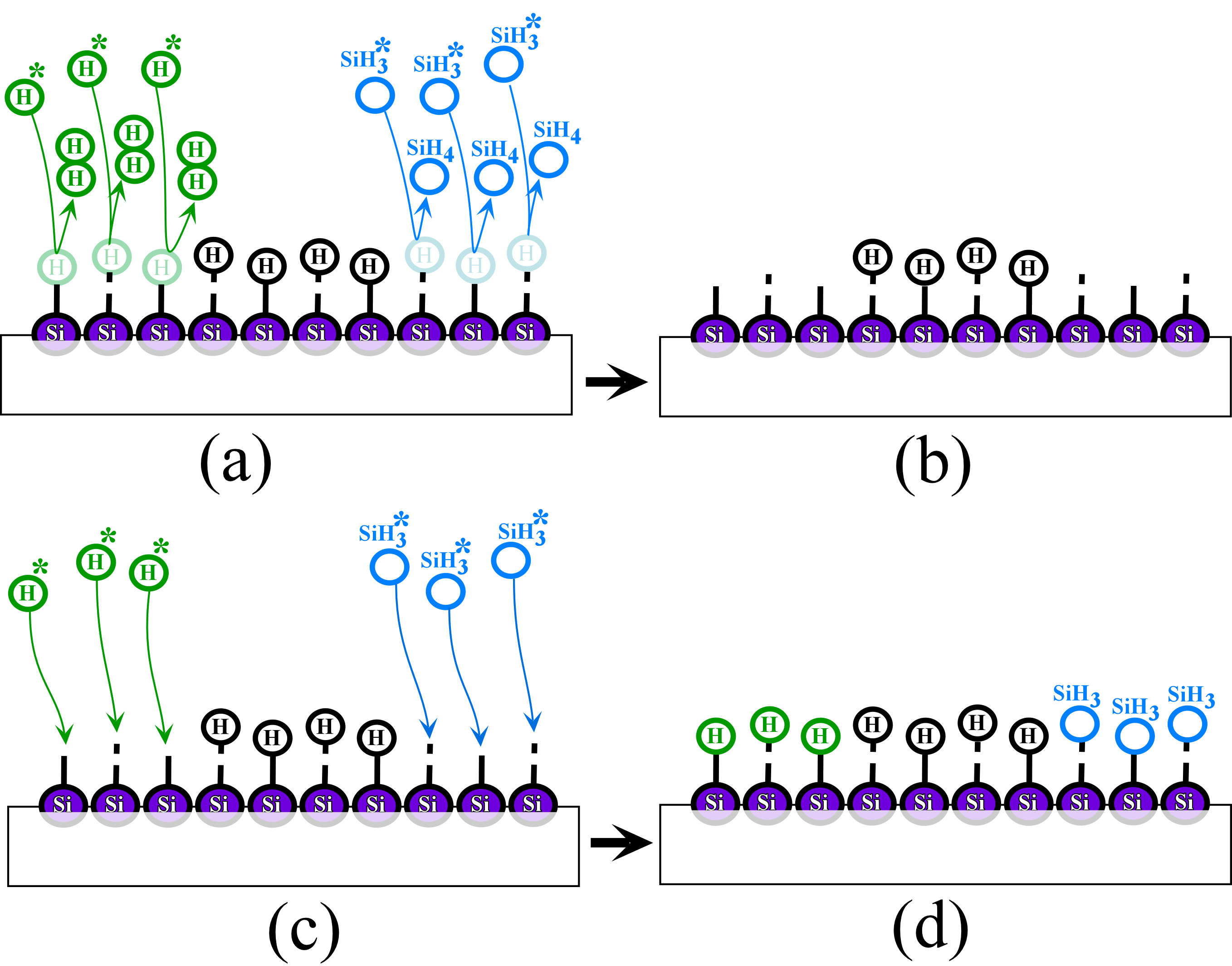} %\label{a} 
	\caption[(a)]{Schematic demonstration of deposition mechanism of thin silicon film on a flat surface }
	\label{fig:fig6}
\end{figure}

%Fig.6.  

\begin{figure}
	\centering
	\includegraphics[width=12cm]{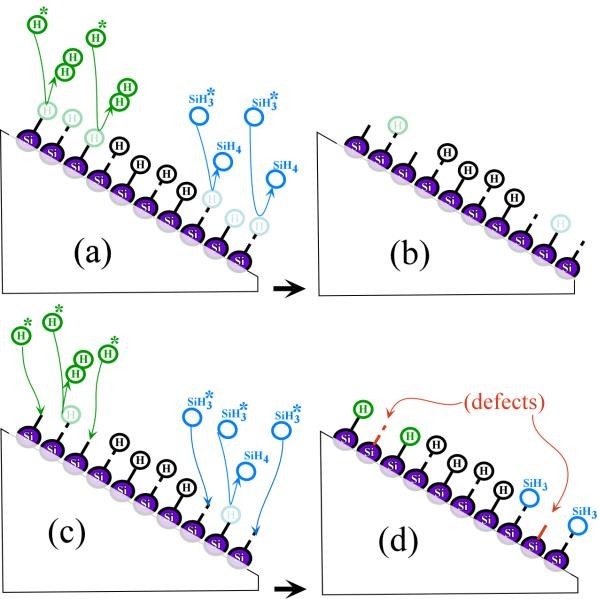} %\label{a} 
	\caption[(a)]{Schematic demonstration of deposition mechanism of thin silicon film on a textured surface }
	\label{fig:fig7}
\end{figure}

%Fig.7.  

\section{Deposition Kinetics}

Based on the above investigation it can be concluded that the deposition of the thin film silicon layers on a textured surface may lead to an increase in the density of electronic defects. For a textured surface, the available surface area is larger than that of a flat surface. Flat surface remains perpendicular to the incident flux of plasma radicals. Generally the $SiH_{3}$ radicals are considered favorable for deposition of a high quality thin silicon film. On a flat surface a large number of the $SiH_{3}$ radicals are expected to arrive as a result the film quality is expected to be higher [20]. This is schematically demonstrated in Fig. 6. The growth surface is usually covered by hydrogen forming Si-H bonds at the surface, Fig. 6(a). Some of these H can be removed by incoming radicals from plasma, Fig. 6(b) and then further incoming radicals (Fig. 6.(c)) can form new bonds on these sites thereby deposits one layer of atom, Fig. 6(d).

On the other hand, the surface of local microscopic structures of a textured surface remains inclined, that is, not perpendicular to the flux of incident radicals, Fig. 7(a), which means the average flux density of incoming plasma radicals to the film growing surface is lower. As a result, rate of removal of surface H becomes lower and micro-void with clustered H may form, Fig. 7(b). It may also be possible that during the film growth, there can be an imbalance between the breaking of Si-H at the surface and film growth, Fig. 7(c), which may lead to formation of increased number of dangling bonds, Fig. 7(d). As a result of the flux density of the plasma radicals that remain lower, some of the weaker or defective bonds remain and gets buried into the film. In this way, plasma deposition of thin silicon film can be more defective for a textured surface than that on a flat surface.

Our investigation of the Cell-A and Cell-B indicates that the active layer of the Cell-A, that was prepared on a textured surface, had more electronic defect ($N_{trap}$) than that of the Cell-B, that was prepared on a flat surface.

Surface texturing s generally adapted for enhanced light trapping and coupling of incident light into the active layer of a solar cell. Ideally this should raise the current density and PCE. But it may also be possible that in the process the fabricated layers of the device will become more defective, therefore, device performance is not as good as expected.

The two cells (Cell-A and Cell-B) are expected to have similarity except the surface texturing. And it can be seen that the device on the textured surface does not perform better than that prepared on a flat surface.

In order to introduce the effect of improved light coupling into the active layer of the Cell-B, its p-type doped window layer is made thinner, from 15 nm to 3 nm. The thinner layer will transmit more light than the thicker one. The estimated characteristics of the device shows that its $V_{oc}$=808.1 mV, $J_{sc}$ = 20.96 mA/cm$^{2}$, FF = 72.73$\%$, and PCE = 12.32$\%$. This is a significant enhancement in device performance. Therefore, it can be concluded that a better device can be fabricated on a non-textured surface than that on a textured one.

\section{Conclusions}
We investigated two different p-i-n type thin film silicon solar cells by numerical analysis. The electronic parameters of the layers were accepted when the J-V characteristic curves of the real and the simulated cells match closely. The primary differences in these two cells were thickness of the active layer and nature of surface texturing. It shows that thin film deposited on a textured surface may acquire an increased defect density than that deposited on a flat surface. Optimized device structure and maximum device performance primarily depends upon the electronic parameters of the active layer like defect density, carrier mobility etc. Optical absorption in the front window layer is considered a loss of light. It can be reduced in various different ways, like using wide band gap material as well as thinner p-type layer, where the latter approach is easily attainable and more effective in improving performance of the device. Generally, the determining parameter for the thickness of the p-layer should be its surface roughness. So a better device performance can be obtained with a thinner p-layer that has very low surface roughness. Furthermore, if the effect of texture induced defects supersedes the effect of light trapping, then using flat surface is preferable for device fabrication.

\end{document}